\documentclass{article}

\PassOptionsToPackage{numbers, compress}{natbib}

\usepackage[preprint]{neurips_2026}

\usepackage[utf8]{inputenc}
\usepackage[T1]{fontenc}
\usepackage{hyperref}
\usepackage{url}
\usepackage{amsfonts}
\usepackage{amsmath}
\usepackage{microtype}
\usepackage[table]{xcolor}
\usepackage{booktabs}
\usepackage{tabularx}
\usepackage{graphicx}

\definecolor{myblue}{HTML}{3A5870}
\definecolor{mygreen}{HTML}{596A42}
\definecolor{myred}{HTML}{995040}

\definecolor{noActionColor}{HTML}{E8F1FB}
\definecolor{internalActionColor}{HTML}{FFF0DB}
\definecolor{interventionColor}{HTML}{EDF2E7}

\title{Rethinking World Models for Biosignals}

\author{%
  \textbf{Xianglong Bao}$^{1,*}$ \quad \textbf{Bosong Huang}$^{1,*}$ \quad
  \textbf{Xiaoda Wang}$^{2,3}$ \quad \textbf{Keqi Han}$^{3}$ \\[0.2em]
  \textbf{Carl Yang}$^{3}$ \quad \textbf{Ming Jin}$^{1,\dagger}$ \quad
  \textbf{Shirui Pan}$^{1,\dagger}$ \\[0.6em]
  $^{1}$Griffith University \quad $^{2}$University of California, Los Angeles \quad
  $^{3}$Emory University \\[0.4em]
  {\small $^{*}$Equal contribution. \quad $^{\dagger}$Corresponding authors.} \\[0.2em]
  {\small \texttt{xianglong.bao@griffithuni.edu.au, mingjinedu@gmail.com, s.pan@griffith.edu.au}}
}

\begin{document}

\maketitle

\begin{abstract}
World models promise to learn how systems evolve and respond to actions, enabling
simulation, planning and decision-making. Bringing this promise to biosignals requires
more than accurate prediction of physiological recordings. We examine existing biosignal
world models through two requirements for intervention use: a genuine action and evidence
of its causal effect. We identify three fundamental bottlenecks rooted in physiology:
\emph{Temporal Latency}, the variable delay between an action and its observable effect;
\emph{System Openness}, the influence of drivers beyond the modelled intervention; and
\emph{Endogenous Dynamics}, the system's ongoing evolution without that intervention.
Drawing on published evidence, we outline corresponding research directions and propose
an \emph{Intervention-Effect Validation Checklist} to guide evaluation. The aim is to
help the biosignal community realise the core promise of world models by learning the
latent mechanisms that govern physiological evolution and responses to interventions.
\end{abstract}

\section{Introduction}

World models learn representations of an environment and its dynamics, allowing an
agent to simulate possible futures before acting \citep{ha2018world,hafner2025mastering}.
The learned dynamics can be reused for planning, policy learning and simulation across
downstream tasks. The broader promise is to learn mechanisms that explain how a system
evolves and responds to actions, supporting decisions through simulation of possible outcomes.

Applications span games \citep{micheli2023iris}, robotics \citep{wu2022daydreamer},
autonomous driving \citep{hu2023gaia} and interactive environments \citep{bruce2024genie}.
In physiology and health, the prospect is to simulate individual trajectories under
alternative interventions \citep{qazi2025beyond,liu2026medical}, with work on clinical
records \citep{wang2026chrono,xu2025meddreamer} and continuously recorded biosignals,
including EEG, ECG and wearable streams
\citep{yin2026eegwm,chen2026ecgwm,luo2026normwear}.

\begin{figure}[t]
  \centering
  \includegraphics[width=\linewidth]{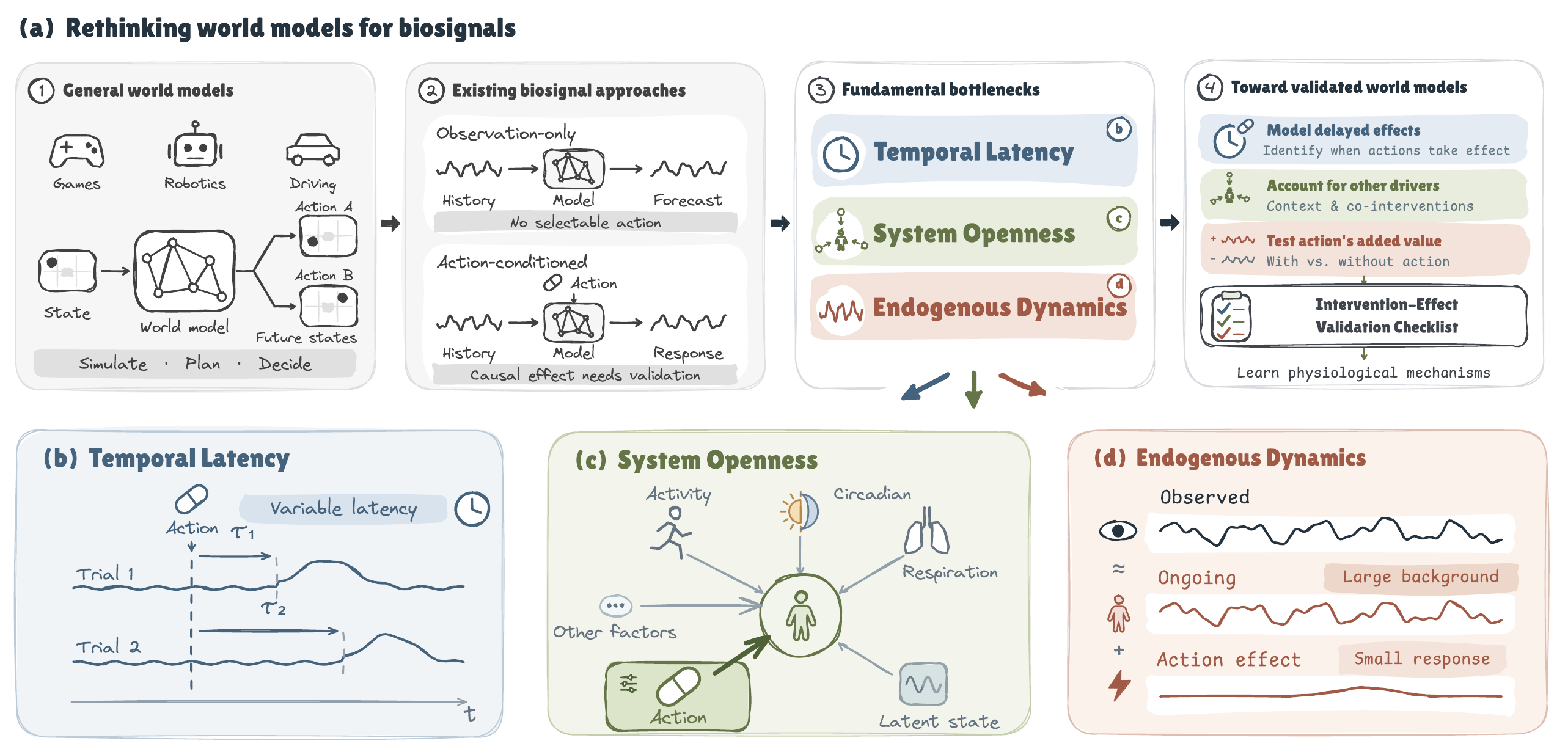}
  \caption{Biosignal world models and three fundamental bottlenecks. (a)~From general
  world models and existing biosignal approaches to bottlenecks and validation.
  (b)~\textcolor{myblue}{\textbf{Temporal Latency}}: the delay between an action and its effect varies between trials.
  (c)~\textcolor{mygreen}{\textbf{System Openness}}
: the action is one of many drivers of physiological state.
  (d)~\textcolor{myred}{\textbf{Endogenous Dynamics}}: ongoing dynamics can exceed the response to the action
  (Section~\ref{sec:bottlenecks}).}
  \label{fig:overview}
\end{figure}

Existing approaches differ in what they model, what counts as an action and how they
are evaluated. We examine whether their evidence supports intervention simulation
(Section~\ref{sec:criterion}), then identify three fundamental bottlenecks:
\textbf{Temporal Latency}, \textbf{System Openness} and \textbf{Endogenous Dynamics}.
Intervention effects can be delayed, other factors also drive the signal, and physiological
activity continues without intervention (Figure~\ref{fig:overview}b--d).
These bottlenecks motivate corresponding research directions and an Intervention-Effect
Validation Checklist, with the goal of learning latent physiological mechanisms.

\section{The state of biosignal world models}
\label{sec:criterion}

\subsection{Requirements for intervention simulation}

Definitions of world models remain unsettled \citep{ding2025survey}. We focus on their
use for intervention simulation and decisions. An action-conditioned model predicts
$p(s_{t+1}\mid s_t,a_t)$, where $s_t$ represents the system's state and $a_t$ an action.
Biosignals only partially observe physiological state, making latent-state learning part
of the problem. Two requirements distinguish this intended use from ordinary prediction.

First, $a_t$ must be a genuine action that could be set differently, such as a stimulus,
drug dose or ventilator setting \citep{sutton2018reinforcement}. Another physiological
channel or disease onset does not itself specify a selectable intervention.

Second, the model must learn the action's causal effect. Ordinary conditional forecasting
has the same mathematical form but may only capture associations. Comparing interventions
requires predicting what happens when an action is set differently
\citep{pearl2009causality,scholkopf2021toward}. Treatment given to sicker patients, for example,
can correlate with worse outcomes even when beneficial. Predicting this association does not
recover its effect. Thus, conditioning forecasts on future inputs or interventions
\citep{lim2021tft,liu2026drift} alone does not establish causal validity. Observational
counterfactual methods require identification assumptions, including consistency, treatment
overlap and no unmeasured confounding given observed history \citep{bica2020crn,seedat2022tecde}.

\subsection{Actions and evidence in existing biosignal models}

Table~\ref{tab:biosignal-models} summarises representative approaches. Physiological context, signal
history and disease transitions do not specify selectable interventions, although their
representations can be useful \citep{weimann2025jepa,li2026eegjepa}.

\begin{table}[!htbp]
\centering
\caption{Representative approaches presented as world models for biosignals and physiology, grouped by action input.}
\label{tab:biosignal-models}
\footnotesize
\setlength{\tabcolsep}{4pt}
\renewcommand{\arraystretch}{1.1}
\begin{tabularx}{\linewidth}{@{}>{\raggedright\arraybackslash}p{0.22\linewidth}
>{\raggedright\arraybackslash}p{0.17\linewidth}
>{\raggedright\arraybackslash}X
>{\raggedright\arraybackslash}p{0.28\linewidth}@{}}
\toprule
\textbf{Model} & \textbf{Signals} & \textbf{Action input} & \textbf{Reported evaluation} \\
\midrule
\rowcolor{noActionColor}
\textbf{Multimodal EEG WM} \citep{yin2026eegwm} & EEG, ECG, EMG, movement & No explicit intervention; multimodal context & Seizure detection \\
\rowcolor{noActionColor}
\textbf{Sonata} \citep{delaney2026sonata} & Trunk inertial signals & No intervention input in reported predictor & Clinical discrimination, fall risk, cohort transfer \\
\midrule
\rowcolor{internalActionColor}
\textbf{Cardiac JEPA} \citep{fernandes2026cardiac} & ECG & Changes in diagnosis labels & Triage, monitoring, sample efficiency \\
\rowcolor{internalActionColor}
\textbf{WM augmentation} \citep{facury2026world} & ECG & Changes in ICD labels & Latent augmentation for multi-label classification \\
\midrule
\rowcolor{interventionColor}
\textbf{NeuroWorld} \citep{dong2026neuroworld} & fMRI & Multimodal movie stimuli & Multi-step prediction, stimulus ablations \\
\rowcolor{interventionColor}
\textbf{ECG-WM} \citep{chen2026ecgwm} & ECG & Drug intervention tokens & Waveform fidelity, risk calibration, drug response \\
\rowcolor{interventionColor}
\textbf{NormWear-2} \citep{luo2026normwear} & Physiological streams & Anaesthesia, insulin, ultrafiltration settings & Forecasting, qualitative action sensitivity \\
\rowcolor{interventionColor}
\textbf{Diffusion-Pref} \citep{zhang2026diffusionpref} & Glucose trajectories & Insulin dose; recorded carbohydrate intake & Retrospective prediction and policy evaluation \\
\rowcolor{interventionColor}
\textbf{HealthFormer} \citep{lutsker2026healthformer} & Multimodal health measures & Diet and medication changes & Nutrition trial; published trial comparisons \\
\bottomrule
\end{tabularx}
\end{table}

Evidence varies across models. Diffusion-Pref reports retrospective evaluation, whereas
HealthFormer compares predictions with a held-out nutrition trial and published randomised
trial outcomes. Such evidence goes beyond action sensitivity, but its scope and assumptions
still matter when interpreting intervention effects. Reviews identify broader validation gaps
\citep{qazi2025beyond,liu2026medical,chen2026trustworthy}. Complementing these empirical
models, \citet{zhang2026physiological} propose an event-conditioned physiological world-model
framework with capability levels and validation protocols.

\section{Three fundamental bottlenecks}
\label{sec:bottlenecks}

\noindent \textbf{Temporal Latency} is the delay between an action and its observable effect. The delay
can vary between subjects and occasions, so aligning an action with a subsequent recording
does not establish when its effect begins. For example, \citet{sheiner1979simultaneous}
modelled the delay between plasma concentration and the paralytic effect of d-tubocurarine
using an effect compartment, with equilibration rates varying across subjects. Observation timing can itself be informative:
deteriorating patients may be measured more often, introducing bias if this timing is
ignored \citep{pullenayegum2023irregular}.

Precisely timestamped stimulation or machine settings resolve uncertainty in action
recording but not in physiological response delay. Evaluation must therefore distinguish
the two and test sensitivity to plausible delays and the observation intervals present
in the data. When several actions precede an observable response, attributing it to the
latest action may be misleading. A world model needs enough action history to account
for effects that emerge after subsequent decisions.

\noindent \textbf{System Openness} means that drivers beyond the modelled intervention influence the
recorded state. For example, \citet{sipinkova1997respiration} found that posture and breathing
frequency changed the spectral properties of heart rate variability in healthy subjects.
Such factors may be only partially observed. Moreover, clinical covariates can predict
both subsequent treatment and outcomes while being affected by earlier treatment,
complicating causal estimation \citep{robins2000marginal}.

Consider a clinician adjusting treatment in response to a deteriorating signal. The
subsequent trajectory may reflect the treatment, the condition that prompted it and other
concurrent changes. If relevant context is unrecorded, conditioning on the treatment does
not separate these contributions \citep{pearl2009causality}. Other drivers can also alter
the signal even when actions are assigned independently. Evaluation must therefore account
for both their contribution to observed changes and their possible influence on action
assignment.

\noindent \textbf{Endogenous Dynamics} are the system's ongoing changes without the modelled intervention.
System Openness concerns other drivers; Endogenous Dynamics concerns the system's own
evolution and how its pre-action state shapes the response. \citet{fox2006coherent}
showed that spontaneous activity explains a significant fraction of trial-to-trial variability
in task-evoked fMRI BOLD responses. Under direct electrical stimulation, \citet{rabuffo2026prestimulus}
analysed simultaneous intracranial and high-density scalp recordings from 36 people with
epilepsy across roughly 320 sessions and more than 10{,}000 single-pulse stimulations.
They found that pre-stimulus brain state predicts responses to identical stimuli, with
whole-brain measures more informative than local ones. Restricting stimulation to trials
with favourable pre-stimulus signatures reduced response variability, suggesting that
ongoing state can be used rather than left as an unknown.

A model can therefore forecast accurately by learning ongoing dynamics while using little
action information. Conversely, a state-dependent response can reflect a genuine interaction
between ongoing activity and intervention. Evaluation must account for pre-action state
when measuring the action's contribution, rather than assume an independent additive effect.

\section{Toward validated biosignal world models}
\label{sec:discussion}

\subsection{Addressing the three bottlenecks}

\textbf{Temporal Latency.} Model action histories and delays on time scales distinct from
signal sampling. Continuous-time formulations offer a starting point for irregular observations
\citep{seedat2022tecde}. Test sensitivity to delays and observation schedules, distinguishing
recorded action timing from physiological effect onset.

\textbf{System Openness.} Record co-interventions, physiological context and measurement
conditions. Controlled action variation can help separate intervention effects from treatment
selection. Observational analyses require identification assumptions and sensitivity tests for
unobserved drivers; transfer across action-assignment policies alone cannot rule out confounding.

\textbf{Endogenous Dynamics.} Compare models with and without action inputs using matched
histories and prediction horizons. Test how responses depend on pre-action state, allowing
interactions between ongoing activity and intervention rather than assuming additive effects.
The pre-stimulus findings of \citet{rabuffo2026prestimulus} motivate this direction.

These are research directions, not established solutions. Progress need not require
larger architectures: switched-linear models with roughly 300\,ms of causal history
explain evoked dynamics under precisely timed intracranial stimulation
\citep{acharya2024predictive}. Well-characterised interventions provide a starting point
for validating more general models.

\subsection{Intervention-Effect Validation Checklist}
\label{sec:intervention-checklist}

Biosignal world models should be evaluated on whether they capture intervention effects
despite Temporal Latency, System Openness and Endogenous Dynamics. The following checklist
translates these three bottlenecks into concrete evaluation questions.

\noindent \textbf{1. Temporal Latency.} Does the world model predict when an intervention's
effect appears in the signal? Do its predictions remain reliable when the delay varies
or observations are taken at irregular intervals? Report how uncertainty in timing
affects the predicted response.

\noindent \textbf{2. System Openness.} Can the world model distinguish intervention effects
from changes caused by other factors, including the condition that prompted treatment?
Compare predicted effects with controlled intervention results where available.
Otherwise, explain the assumptions needed to attribute changes to the intervention
and assess how unrecorded factors could affect the conclusions.

\noindent \textbf{3. Endogenous Dynamics.} Does providing the action improve predictions
beyond what the model learns from the signal's history alone? Compare both using
the same past observations and predicting equally far ahead. Does the model capture
how the response changes with the state before intervention? Report the improvement
and its uncertainty, including cases where the action adds little.

\subsection{Learning latent physiological mechanisms}

Addressing these bottlenecks and validating intervention effects can bring biosignal world
models closer to learning latent physiological mechanisms. Representations must capture
delayed action effects, relevant context and ongoing state to support simulation across
intervention settings. Forecasting and representation learning can contribute, but latent
representations alone do not identify physiological mechanisms. Identifiability of latent
states and controlled dynamics depends on system assumptions and action variation in the
data \citep{zhang2026identifiability}.

\section{Conclusion}

World models promise to learn how systems evolve and respond to actions. Realising this
promise for biosignals requires confronting Temporal Latency, System Openness and
Endogenous Dynamics, even when forecasting is accurate. The research directions and Intervention-Effect Validation Checklist proposed
here connect these bottlenecks to model development and evaluation. Progress should be
judged by how reliably models capture physiological responses to interventions, bringing
the field closer to learning the latent mechanisms that govern those responses.

\clearpage
\bibliographystyle{plainnat}
\bibliography{references}

\typeout{get arXiv to do 4 passes: Label(s) may have changed. Rerun}
\end{document}